\documentclass[11pt]{article}           
\usepackage{amsmath}

\newtheorem{Theorem}{Theorem}

\newtheorem{Remark}{Remark}
\newtheorem{Lemma}{Lemma}

\usepackage{graphics}

\date{}

\title{
The Bargmann symmetry constraint and binary nonlinearization of the
super Dirac systems}
\author{Jing Yu$^{a,b}$, Jingsong He$^{b, c}$\footnote{Corresponding author, E-mail
address:jshe@ustc.edu.cn, hejingsong@nbu.edu.cn} ,  Wen-Xiu
Ma$^{d}$,  Yi Cheng$^{b}$
\vspace{4mm}\\
$^{a}$School of Science, Hangzhou Dianzi University, Hangzhou,
Zhejiang,
310018, P. R. China\\
$^{b}$Department of Mathematics, University of Science and
Technology of China, Hefei, Anhui,\\ 230026,  China\\
$^{c}$Department of Mathematics, Ningbo University, Ningbo,
Zhejiang, 315211,
 China\\
$^{d}$Department of Mathematics and Statistics, University of South Florida,\\
Tampa, FL 33620-5700, USA}
\begin{document}
 \maketitle


\begin{abstract}
An explicit Bargmann symmetry constraint is computed and its
 associated binary nonlinearization of Lax pairs is
carried out for the super Dirac systems. Under the obtained symmetry
constraint, the n-th flow of the super Dirac hierarchy is decomposed
into two super finite-dimensional integrable Hamiltonian systems,
defined over the supersymmetry manifold $R^{4N|2N}$ with the
corresponding dynamical variables $x$ and $t_n$. The integrals of
motion required for Liouville integrability are explicitly given.


\noindent{\bf PACS codes(2008):}\ 02.30.Ik, 02.90.+p,
\end{abstract}

\section{Introduction}
For almost twenty years, much attention has been paid to the
construction of finite-dimensional integrable systems from soliton
equations by using symmetry constraints. Either (2+1)-dimensional
soliton equations \cite{KSS,CL,CY} or (1+1)-dimensional soliton
equations \cite{MS,Ma1995}
 can be decomposed
into compatible finite-dimensional integrable systems. It is known
that a crucial idea in carrying out symmetry constraints is the
nonlinearization of Lax pairs for soliton hierarchies, and symmetry
constraints give relations of potentials with eigenfunctions and
adjoint eigenfunctions of Lax pairs so that solutions to soliton
equations can be obtained by solving Jacobi inversion problems
\cite{MZ}. The nonlinearization of Lax pairs is classified into
mono-nonlinearization \cite{ZengLi,CCW,CG} and binary
nonlinearization  \cite{M,MaF1996,MaZhou}.

The technique of nonlinearization has been successfully applied to
many well-known (1+1)-dimensional soliton equations, such as the
AKNS system \cite{MS}\cite{CCW}, the KdV equation \cite{Ma1995} and
the Dirac system \cite{Ma1}. But there are few results on
nonlinearization of super integrable systems, existing in the
literature. Studies provides many examples of supersymmetry
integrable systems, with super dependent variables and/or super
independent variables \cite{pop,mo,lm,kup,Lz1+2,lz3}. Very recently,
nonlinearization was made for the super AKNS system \cite{HYZC} and
the corresponding super finite-dimensional Hamiltonian systems were
generated. In this paper, we would like to analyze binary
nonlinearization for the super Dirac systems under a Bargmann
symmetry constraint.

The paper is organized as follows. In the next section, we will
recall the super Dirac soliton hierarchy and its super Hamiltonian
structure. Then in section 3, we compute a Bargmann symmetry
constraint for the potential of the super Dirac hierarchy. In
section 4, we apply binary nonlinearization to the super Dirac
hierarchy, and then obtain super finite-dimensional integrable
Hamiltonian systems on the supersymmetry manifold $R^{4N|2N}$, whose
integrals of motion are explicitly given. Some conclusions and
remarks are listed in section 5.

\section{The Super Dirac Hierarchy}

The super Dirac spectral problem associated with the Lie
super-algebra  $B(0, 1)$ is given by \cite{MaHQ-preprint2007}
\begin{equation}\label{D1}
\phi_x=U\phi,\quad U=\left(\begin{array}{ccc} r&\lambda+s&\alpha\\
-\lambda+s&-r&\beta\\\beta&-\alpha&0\end{array}\right),\quad
u=\left(\begin{array}{c}r\\s\\\alpha\\\beta\end{array}\right),\quad
\phi=\left(\begin{array}{c}\phi_1\\\phi_2\\\phi_3\end{array}\right),\end{equation}
where $\lambda$ is a spectral parameter, $r$ and $s$ are even
variables, and  $\alpha$ and $\beta$ are odd variables.
Taking
$$V=\left(\begin{array}{ccc}
C&A+B&\rho\\A-B&-C&\delta\\\delta&-\rho&0\end{array}\right),$$ the
co-adjoint equation associated with (\ref{D1}) $V_x=[U, V]$ gives
\begin{equation}\label{D4}\left\{\begin{array}{l}
A_x=-2\lambda C+2rB-\alpha\rho+\beta\delta,\\
B_x=2rA-2sC-\alpha\rho-\beta\delta,\\
C_x=2\lambda A-2sB+\alpha\delta+\beta\rho,\\
\rho_x=-\beta(A+B)-\alpha C+(\lambda+s)\delta+r\rho,\\
\delta_x=(-\lambda+s)\rho-r\delta-\alpha(A-B)+\beta
C.\end{array}\right.\end{equation} If we set
\begin{equation}\label{D4`}A=\sum_{i\geq0}A_i\lambda^{-i},
B=\sum_{i\geq0}B_i\lambda^{-i}, C=\sum_{i\geq0}C_i\lambda^{-i},
\rho=\sum_{i\geq0}\rho_i\lambda^{-i},
\delta=\sum_{i\geq0}\delta_i\lambda^{-i},\end{equation} equation
(\ref{D4}) is equivalent to
\begin{equation}\label{D5}\left\{\begin{array}{l}
A_0=C_0=\rho_0=\delta_0=0,\\
A_{i+1}=\frac{1}{2}C_{i,
x}+sB_i-\frac{1}{2}\alpha\delta_i-\frac{1}{2}\beta\rho_i,\quad
i\geq0,\\
C_{i+1}=-\frac{1}{2}A_{i,
x}+rB_i-\frac{1}{2}\alpha\rho_i+\frac{1}{2}\beta\delta_i,\quad
i\geq0,\\
\rho_{i+1}=-\delta_{i, x}-r\delta_i+s\rho_i-\alpha(A_i-B_i)+\beta
C_i,\quad i\geq0,\\
\delta_{i+1}=\rho_{i, x}-r\rho_i-s\delta_i+\beta(A_i+B_i)+\alpha
C_i,\quad i\geq0,\\
B_{i+1,
x}=2rA_{i+1}-2sC_{i+1}-\alpha\rho_{i+1}-\beta\delta_{i+1},\quad
i\geq0,\end{array}\right.\end{equation} which results in recurrence
relations
\begin{equation}\label{D6}\left\{\begin{array}{l}
(C_{i+1}, A_{i+1}, \delta_{i+1}, -\rho_{i+1})^{T}={\cal L}(C_i, A_i,
\delta_i, -\rho_i)^{T},\quad i\geq0,\\
B_{i}=\partial^{-1}(2rA_i-2sC_i-\alpha\rho_i-\beta\delta_i),\quad
i\geq0,
\end{array}\right.\end{equation}
where
$${\cal L}=\left(\begin{array}{cccc}
-2r\partial^{-1}s&-\frac{1}{2}\partial+2r\partial^{-1}r&
\frac{1}{2}\beta-r\partial^{-1}\beta&\frac{1}{2}\alpha+r\partial^{-1}\alpha\\
\frac{1}{2}\partial-2s\partial^{-1}s&2s\partial^{-1}r&
-\frac{1}{2}\alpha-s\partial^{-1}\beta&\frac{1}{2}\beta+s\partial^{-1}\alpha\\
\alpha-2\beta\partial^{-1}s&\beta+2\beta\partial^{-1}r&
-s-\beta\partial^{-1}\beta&-\partial+r+\beta\partial^{-1}\alpha\\
-\beta+2\alpha\partial^{-1}s&\alpha-2\alpha\partial^{-1}r&
\partial+r+\alpha\partial^{-1}\beta&s-\alpha\partial^{-1}\alpha
\end{array}\right).$$
Upon choosing the initial conditions
$$A_0=C_0=\rho_0=\delta_0=0,\quad B_0=1,$$
all other $A_i, B_i, C_i, \rho_i, \delta_i, \ i\geq1,$ can be worked
out uniquely by the recurrence relations (\ref{D6}). The first few
results are as follows
\begin{eqnarray*}A_1&=&s,\ B_1=0,\
C_1=r,\ \rho_1=\alpha,\ \delta_1=\beta,\\
A_2&=&\frac{1}{2}r_x,\ B_2=\frac{1}{2}(r^{2}+s^{2})+\alpha\beta,\
C_2=-\frac{1}{2}s_x,\ \rho_2=-\beta_x,\ \delta_2=\alpha_x,\\
A_3&=&-\frac{1}{4}s_{xx}+\frac{1}{2}(r^{2}+s^{2})s+s\alpha\beta
-\frac{1}{2}\alpha\alpha_x+\frac{1}{2}\beta\beta_x,\\
B_3&=&-\frac{1}{2}(rs_x-r_xs)+\alpha\alpha_x+\beta\beta_x,\\
C_3&=&-\frac{1}{4}r_{xx}+\frac{1}{2}(r^{2}+s^{2})r+r\alpha\beta
+\frac{1}{2}\alpha\beta_x-\frac{1}{2}\alpha_x\beta,\\
\rho_3&=&-\alpha_{xx}+\frac{1}{2}(r^{2}+s^{2})\alpha-\frac{1}{2}r_x\alpha
-\frac{1}{2}s_x\beta-r\alpha_x-s\beta_x,\\
\delta_3&=&-\beta_{xx}+\frac{1}{2}(r^{2}+s^{2})\beta+\frac{1}{2}r_x\beta
-\frac{1}{2}s_x\alpha-s\alpha_x+r\beta_x.\end{eqnarray*}

Let us associate the spectral problem (\ref{D1}) with the following
auxiliary spectral problem
\begin{equation}\label{D7}
\phi_{t_n}=V^{(n)}\phi=(\lambda^{n}V)_+\phi,
\end{equation}
with $$V^{(n)}=\sum_{i=0}^{n}\left(\begin{array}{ccc}
C_i&A_i+B_i&\rho_i\\
A_i-B_i&-C_i&\delta_i\\
\delta_i&-\rho_i&0\end{array}\right)\lambda^{n-i},$$
 where the plus symbol "+" denotes  taking the non-negative part in the power of $\lambda$.

The compatible conditions of the spectral problem (\ref{D1}) and the
auxiliary spectral problem (\ref{D7}) are
\begin{equation}\label{D8}
U_{t_n}-V_x^{(n)}+[U, V^{(n)}]=0, \quad n\geq0,
\end{equation}
 which infer the super Dirac soliton hierarchy
\begin{equation}\label{D9}
u_{t_n}=K_n=(2A_{n+1}, -2C_{n+1}, \delta_{n+1},
-\rho_{n+1})^{T},\quad n\geq0.\end{equation} Here $u_{t_n}=K_n$ in
(\ref{D9}) is called the n-th Dirac flow of the hierarchy.

Using the super trace identity \cite{MaHQ-preprint2007, Hu}
\begin{equation}\label{D10}
\frac{\delta}{\delta u}\int {\rm Str}(V\frac{\partial
U}{\partial\lambda})dx
=(\lambda^{-\gamma}\frac{\partial}{\partial\lambda}\lambda^{\gamma})
Str(\frac{\partial U}{\partial u}V),
\end{equation}
where Str means the super trace, we can have
\begin{equation}\label{D11}
\left(\begin{array}{c}C_{i+1}\\A_{i+1}\\\delta_{i+1}\\-\rho_{i+1}\end{array}\right)
=\frac{\delta}{\delta u}H_i,\quad
H_i=\int\frac{B_{i+2}}{i+1}dx,\quad i\ge 0.\end{equation} Therefore,
the super Dirac soliton hierarchy (\ref{D9}) can be written as the
following super Hamiltonian form:
\begin{equation}\label{D12}
u_{t_n}=J\frac{\delta H_n}{\delta u},\end{equation} where
\[
J=\left(\begin{array}{cccc}
0&2&0&0\\-2&0&0&0\\0&0&1&0\\0&0&0&1\end{array}\right)\] is a
supersymplectic operator, and $H_n$ is given by (\ref{D11}).

The first non-trivial nonlinear equation of hierarchy (\ref{D12}) is
given by the second Dirac flow
\begin{equation}\label{D13}\left\{\begin{array}{l}
r_{t_2}=-\frac{1}{2}s_{xx}+(r^{2}+s^{2})s+2s\alpha\beta-\alpha\alpha_x+\beta\beta_x,\\
s_{t_2}=\frac{1}{2}r_{xx}-(r^{2}+s^{2})r-2r\alpha\beta-\alpha\beta_x+\alpha_x\beta,\\
\alpha_{t_2}=-\beta_{xx}+\frac{1}{2}(r^{2}+s^{2})\beta+r\beta_x-s\alpha_x+\frac{1}{2}r_x\beta
-\frac{1}{2}s_x\alpha,\\
\beta_{t_2}=\alpha_{xx}-\frac{1}{2}(r^{2}+s^{2})\alpha+r\alpha_x+s\beta_x+\frac{1}{2}r_x\alpha
+\frac{1}{2}s_x\beta,\end{array}\right.\end{equation} which
possesses a Lax pair of U defined in (\ref{D1}) and $V^{(2)}$
defined by
$$V^{(2)}=\left(\begin{array}{ccc}
r\lambda-\frac{1}{2}s_x&\lambda^{2}+s\lambda
+\frac{1}{2}r_x+\frac{1}{2}(r^{2}+s^{2})+\alpha\beta&
\alpha\lambda-\beta_x\\
-\lambda^{2}+s\lambda+\frac{1}{2}r_x-\frac{1}{2}(r^{2}+s^{2})-\alpha\beta&
-r\lambda+\frac{1}{2}s_x&\beta\lambda+\alpha_x\\
\beta\lambda+\alpha_x&-\alpha\lambda+\beta_x&0\end{array}\right).$$

\begin{Remark}\label{Rem1}
We consider all differential equations in the real field and explore the Liouville integrability on real symplectic manifolds.
We did not see any equivalence between the real Dirac soliton hierarchy and the real AKNS soliton hierarchy. For example,
it is clear that the AKNS system of nonlinear Schr\"odinger equations and the Dirac system of nonlinear Schr\"odinger equations
can not be transformed into each other by any real linear transformations. There is a similar situation between the super Dirac
 soliton hierarchy and the super AKNS soliton hierarchy, and between the Liouville integrable constrained flows associated with
 the two super soliton hierarchies.

\end{Remark}

\section{The Bargmann symmetry constraint}
In order to compute a Bargmann symmetry constraint, we consider the
following adjoint spectral problem of the spectral problem
(\ref{D1}):
\begin{equation}
\psi_x=-U^{St}\psi=\left(\begin{array}{ccc} -r&\lambda-s&\beta\\
-\lambda-s&r&-\alpha\\-\alpha&-\beta&0\end{array}\right)\psi,\quad
\psi=\left(\begin{array}{c}\psi_1\\\psi_2\\\psi_3\end{array}\right),
\end{equation}
where St means the super transposition. The following result is a
general formula for the variational derivative with respect to the
potential $u$ (see \cite{MS} for the classical case).

\begin{Lemma}\label{Lemma1}
Let $U(u, \lambda)$ be an even matrix of order $m+n$ depending on
$u, u_x, u_{xx}, \cdots$ and a parameter $\lambda$. Suppose that
$\phi=(\phi_e, \phi_o)^{T}$ and $\psi=(\psi_e, \psi_o)^{T}$ satisfy
the spectral problem and the adjoint spectral problem $$\phi_x=U(u,
\lambda)\phi, \quad \psi_x=-U^{St}(u, \lambda)\psi,$$ where
$\phi_e=(\phi_1, \cdots, \phi_m)$ and $\psi_e=(\psi_1, \cdots,
\psi_m)$ are even eigenfunctions, and $\phi_o=(\phi_{m+1}, \cdots,
\phi_{m+n})$ and $\psi_o=(\psi_{m+1}, \cdots, \psi_{m+n})$ are odd
eigenfunctions. Then, the variational derivative of the spectral
parameter $\lambda$ with respect to the potential $u$ is given by
\begin{equation}\label{L1}
\frac{\delta\lambda}{\delta u}=\frac{(\psi_e,
(-1)^{p(u)}\psi_o)(\frac{\partial U}{\partial
u})\phi}{-\int\psi^{T}(\frac{\partial U}{\partial \lambda})\phi
dx},\end{equation} where we denote
\begin{equation}
p(v)=\left\{\begin{array}{l} 0,\quad $$v$ is an even variable$,\\
1,\quad $$v$ is an odd variable$.\end{array}\right.\end{equation}
\end{Lemma}

By Lemma \ref{Lemma1}, it is not difficult to find that
\begin{equation}\label{D15}
\frac{\delta\lambda}{\delta u}=\left(\begin{array}{c}
\psi_1\phi_1-\psi_2\phi_2\\\psi_1\phi_2+\psi_2\phi_1\\
\psi_1\phi_3+\psi_3\phi_2\\\psi_2\phi_3-\psi_3\phi_1\end{array}\right).\end{equation}
When zero boundary conditions
$\lim_{|x|\rightarrow\infty}\phi=\lim_{|x|\rightarrow\infty}\psi=0$
are imposed, we can obtain a characteristic property -- a recurrence
relation for the variational derivative of $\lambda$:
\begin{equation}\label{D16}
{\cal L}\frac{\delta\lambda}{\delta
u}=\lambda\frac{\delta\lambda}{\delta u},\end{equation} where ${\cal
L}$ and $\frac{\delta\lambda}{\delta u}$ are given by (\ref{D6}) and
(\ref{D15}), respectively.

Let us now discuss two spatial and temporal systems:
\begin{equation}\label{D17}\left\{\begin{array}{l}
\left(\begin{array}{c}\phi_{1j}\\\phi_{2j}\\\phi_{3j}\end{array}\right)_x
=U(u,
\lambda_j)\left(\begin{array}{c}\phi_{1j}\\\phi_{2j}\\\phi_{3j}\end{array}\right)
=\left(\begin{array}{ccc}r&\lambda_j+s&\alpha\\
-\lambda_j+s&-r&\beta\\\beta&-\alpha&0\end{array}\right)
\left(\begin{array}{c}\phi_{1j}\\\phi_{2j}\\\phi_{3j}\end{array}\right),\\
\left(\begin{array}{c}\psi_{1j}\\\psi_{2j}\\\psi_{3j}\end{array}\right)_x=
-U^{St}(u,
\lambda_j)\left(\begin{array}{c}\psi_{1j}\\\psi_{2j}\\\psi_{3j}\end{array}\right)
=\left(\begin{array}{ccc}-r&\lambda_j-s&\beta\\-\lambda_j-s&r&-\alpha\\
-\alpha&-\beta&0\end{array}\right)
\left(\begin{array}{c}\psi_{1j}\\\psi_{2j}\\\psi_{3j}\end{array}\right)
;\end{array}\right.\end{equation}
\begin{equation}\label{D18}\left\{\begin{array}{l}
\left(\begin{array}{c}\phi_{1j}\\\phi_{2j}\\\phi_{3j}\end{array}\right)_{t_n}
=V^{(n)}(u,
\lambda_j)\left(\begin{array}{c}\phi_{1j}\\\phi_{2j}\\\phi_{3j}\end{array}\right)\\
\qquad \qquad \quad=\left(\begin{array}{ccc}
\sum\limits_{i=0}^{n}C_i\lambda_j^{n-i}&\sum\limits_{i=0}^{n}(A_i+B_i)\lambda_j^{n-i}&
\sum\limits_{i=0}^{n}\rho_i\lambda_j^{n-i}\\\sum\limits_{i=0}^{n}(A_i-B_i)\lambda_j^{n-i}&
-\sum\limits_{i=0}^{n}C_i\lambda_j^{n-i}&\sum\limits_{i=0}^{n}\delta_i\lambda_j^{n-i}\\
\sum\limits_{i=0}^{n}\delta_i\lambda_j^{n-i}&-\sum\limits_{i=0}^{n}\rho_i\lambda_j^{n-i}&0
\end{array}\right)\left(\begin{array}{c}\phi_{1j}\\\phi_{2j}\\\phi_{3j}\end{array}\right)
,\\
\left(\begin{array}{c}\psi_{1j}\\\psi_{2j}\\\psi_{3j}\end{array}\right)_{t_n}
=-(V^{(n)})^{St}(u,
\lambda_j)\left(\begin{array}{c}\psi_{1j}\\\psi_{2j}\\\psi_{3j}\end{array}\right)\\
\qquad \qquad \quad
=\left(\begin{array}{ccc}-\sum\limits_{i=0}^{n}C_i\lambda_j^{n-i}&
-\sum\limits_{i=0}^{n}(A_i-B_i)\lambda_j^{n-i}&\sum\limits_{i=0}^{n}\delta_i\lambda_j^{n-i}\\
-\sum\limits_{i=0}^{n}(A_i+B_i)\lambda_j^{n-i}&\sum\limits_{i=0}^{n}C_i\lambda_j^{n-i}&
-\sum\limits_{i=0}^{n}\rho_i\lambda_j^{n-i}\\-\sum\limits_{i=0}^{n}\rho_i\lambda_j^{n-i}&
-\sum\limits_{i=0}^{n}\delta_i\lambda_j^{n-i}&0\end{array}\right)
\left(\begin{array}{c}\psi_{1j}\\\psi_{2j}\\\psi_{3j}\end{array}\right);
\end{array}\right.\end{equation}
where $1\leq j\leq N$ and $\lambda_1, \cdots, \lambda_N$ are $N$
distinct spectral parameters. Now for the systems (\ref{D17}) and
(\ref{D18}), we have the following symmetry constraints
\begin{equation}\label{D19}
\frac{\delta}{\delta
u}H_k=\sum_{j=1}^{N}\frac{\delta\lambda_j}{\delta u},\ k\ge
0.\end{equation} The symmetry constraint in the case of $k=0$ is
called a Bargmann symmetry constraint \cite{MaZhou}. It leads to an
explicit expression for the potential $u$, i.e.,
\begin{equation}\label{D20}\left\{\begin{array}{l}
r=<\Psi_1, \Phi_1>-<\Psi_2, \Phi_2>,\\
s=<\Psi_1, \Phi_2>+<\Psi_2, \Phi_1>,\\
\alpha=-<\Psi_2, \Phi_3>+<\Psi_3, \Phi_1>,\\
\beta=<\Psi_1, \Phi_3>+<\Psi_3.
\Phi_2>,\end{array}\right.\end{equation} where we use the following
notation, \[ \Phi_i=(\phi_{i1}, \cdots, \phi_{iN})^{T}, \
\Psi_i=(\psi_{i1}, \cdots, \psi_{iN})^{T},\ i=1, 2, 3,
\]
 and $<\cdot,\cdot>$ denotes
the standard inner product of the Euclidian space $R^{N}$.

\section{Binary nonlinearization }

In this section, we want to perform binary nonlinearization for the
Lax pairs and adjoint Lax pairs of the super Dirac hierarchy
(\ref{D12}). To this end, let us substitute (\ref{D20}) into the Lax
pairs and adjoint Lax pairs (\ref{D17}) and (\ref{D18}), and then we
obtain the following nonlinearized Lax pairs and adjoint Lax pairs
\begin{equation}\label{D21}\left\{\begin{array}{l}
\left(\begin{array}{c}\phi_{1j}\\\phi_{2j}\\\phi_{3j}\end{array}\right)_x
=U(\tilde{u},
\lambda_j)\left(\begin{array}{c}\phi_{1j}\\\phi_{2j}\\\phi_{3j}\end{array}\right)
=\left(\begin{array}{ccc}\tilde{r}&\lambda_j+\tilde{s}&\tilde{\alpha}\\
-\lambda_j+\tilde{s}&-\tilde{r}&\tilde{\beta}\\\tilde{\beta}&-\tilde{\alpha}&0\end{array}\right)
\left(\begin{array}{c}\phi_{1j}\\\phi_{2j}\\\phi_{3j}\end{array}\right),\\
\left(\begin{array}{c}\psi_{1j}\\\psi_{2j}\\\psi_{3j}\end{array}\right)_x=
-U^{St}(\tilde{u},
\lambda_j)\left(\begin{array}{c}\psi_{1j}\\\psi_{2j}\\\psi_{3j}\end{array}\right)
=\left(\begin{array}{ccc}-\tilde{r}&\lambda_j-\tilde{s}&\tilde{\beta}\\
-\lambda_j-\tilde{s}&\tilde{r}&-\tilde{\alpha}\\
-\tilde{\alpha}&-\tilde{\beta}&0\end{array}\right)
\left(\begin{array}{c}\psi_{1j}\\\psi_{2j}\\\psi_{3j}\end{array}\right);
\end{array}\right.\end{equation}
\begin{equation}\label{D22}\left\{\begin{array}{l}
\left(\begin{array}{c}\phi_{1j}\\\phi_{2j}\\\phi_{3j}\end{array}\right)_{t_n}
=V^{(n)}(\tilde{u},
\lambda_j)\left(\begin{array}{c}\phi_{1j}\\\phi_{2j}\\\phi_{3j}\end{array}\right)\\
\qquad \qquad \quad=\left(\begin{array}{ccc}
\sum\limits_{i=0}^{n}\tilde{C}_i\lambda_j^{n-i}&\sum\limits_{i=0}^{n}(\tilde{A}_i+\tilde{B}_i)\lambda_j^{n-i}&
\sum\limits_{i=0}^{n}\tilde{\rho}_i\lambda_j^{n-i}\\\sum\limits_{i=0}^{n}(\tilde{A}_i-\tilde{B}_i)\lambda_j^{n-i}&
-\sum\limits_{i=0}^{n}\tilde{C}_i\lambda_j^{n-i}&\sum\limits_{i=0}^{n}\tilde{\delta}_i\lambda_j^{n-i}\\
\sum\limits_{i=0}^{n}\tilde{\delta}_i\lambda_j^{n-i}&-\sum\limits_{i=0}^{n}\tilde{\rho}_i\lambda_j^{n-i}&0
\end{array}\right)\left(\begin{array}{c}\phi_{1j}\\\phi_{2j}\\\phi_{3j}\end{array}\right)
,\\
\left(\begin{array}{c}\psi_{1j}\\\psi_{2j}\\\psi_{3j}\end{array}\right)_{t_n}
=-(V^{(n)})^{St}(\tilde{u},
\lambda_j)\left(\begin{array}{c}\psi_{1j}\\\psi_{2j}\\\psi_{3j}\end{array}\right)\\
\qquad \qquad \quad
=\left(\begin{array}{ccc}-\sum\limits_{i=0}^{n}\tilde{C}_i\lambda_j^{n-i}&
-\sum\limits_{i=0}^{n}(\tilde{A}_i-\tilde{B}_i)\lambda_j^{n-i}&\sum\limits_{i=0}^{n}\tilde{\delta}_i\lambda_j^{n-i}\\
-\sum\limits_{i=0}^{n}(\tilde{A}_i+\tilde{B}_i)\lambda_j^{n-i}&\sum\limits_{i=0}^{n}\tilde{C}_i\lambda_j^{n-i}&
-\sum\limits_{i=0}^{n}\tilde{\rho}_i\lambda_j^{n-i}\\-\sum\limits_{i=0}^{n}\tilde{\rho}_i\lambda_j^{n-i}&
-\sum\limits_{i=0}^{n}\tilde{\delta}_i\lambda_j^{n-i}&0\end{array}\right)
\left(\begin{array}{c}\psi_{1j}\\\psi_{2j}\\\psi_{3j}\end{array}\right),
\end{array}\right.\end{equation}
where $1\leq j\leq N$ and $\tilde{P}$ means an expression of $P(u)$
under the explicit constraint (\ref{D20}). Note that the spatial
part of the nonlinearized system (\ref{D21}) is a system of ordinary
differential equations with an independent variable $x$, but for a
given $n$ ($n\geq2$), the $t_n$-part of the nonlinearized system
(\ref{D22}) is a system of ordinary differential equations.
Obviously, the system (\ref{D21}) can be written as
\begin{equation}\label{D23}\left\{\begin{array}{l}
\Phi_{1, x}=(<\Psi_1, \Phi_1>-<\Psi_2,
\Phi_2>)\Phi_1+(\Lambda+<\Psi_1, \Phi_2>+<\Psi_2,
\Phi_1>)\Phi_2+(-<\Psi_2, \Phi_3>\\\qquad \quad+<\Psi_3, \Phi_1>)\Phi_3,\\
\Phi_{2, x}=(-\Lambda+<\Psi_1, \Phi_2>+<\Psi_2,
\Phi_1>)\Phi_1-(<\Psi_1, \Phi_1>-<\Psi_2, \Phi_2>)\Phi_2+(<\Psi_1,
\Phi_3>\\\qquad \quad+<\Psi_3, \Phi_2>)\Phi_3,\\
\Phi_{3, x}=(<\Psi_1, \Phi_3>+<\Psi_3, \Phi_2>)\Phi_1-(-<\Psi_2,
\Phi_3>+<\Psi_3, \Phi_1>)\Phi_2,\\
\Psi_{1, x}=-(<\Psi_1, \Phi_1>-<\Psi_2,
\Phi_2>)\Psi_1+(\Lambda-<\Psi_1, \Phi_2>-<\Psi_2, \Phi_1>)\Psi_1
+(<\Psi_1, \Phi_3>\\\qquad\quad+<\Psi_3, \Phi_2>)\Psi_3,\\
\Psi_{2, x}=-(\Lambda+<\Psi_1, \Phi_2>+<\Psi_2,
\Phi_1>)\Psi_1+(<\Psi_1, \Phi_1>-<\Psi_2, \Phi_2>)\Psi_2-(-<\Psi_2,
\Phi_3>\\\qquad \quad+<\Psi_3, \Phi_1>)\Psi_3,\\
\Psi_{3, x}=-(-<\Psi_2, \Phi_3>+<\Psi_3, \Phi_1>)\Psi_1-(<\Psi_1,
\Phi_3>+<\Psi_3, \Phi_2>)\Psi_2,\end{array}\right.
\end{equation}
where $\Lambda=diag(\lambda_1, \cdots, \lambda_N).$ When $n=1$, the
system (\ref{D22}) is exactly the system (\ref{D21}) with $t_1=x$.
When $n=2$, the system (\ref{D22}) is
\begin{equation}\label{D24}\left\{\begin{array}{l}
\Phi_{1, t_2}=(\tilde{r}\Lambda-\frac{1}{2}\tilde{s}_x)\Phi_1
+(\Lambda^{2}+\tilde{s}\Lambda+\frac{1}{2}\tilde{r}_x+\frac{1}{2}(\tilde{r}^{2}+\tilde{s}^{2})
+\tilde{\alpha}\tilde{\beta})\Phi_2+(\tilde{\alpha}\Lambda-\tilde{\beta}_x)\Phi_3,\\
\Phi_{2,
t_2}=(-\Lambda^{2}+\tilde{s}\Lambda+\frac{1}{2}\tilde{r}_x-\frac{1}{2}(\tilde{r}^{2}+\tilde{s}^{2})
-\tilde{\alpha}\tilde{\beta})\Phi_1-(\tilde{r}\Lambda-\frac{1}{2}\tilde{s}_x)\Phi_2
+(\tilde{\beta}\Lambda+\tilde{\alpha}_x)\Phi_3,\\
\Phi_{3,
t_2}=(\tilde{\beta}\Lambda+\tilde{\alpha}_x)\Phi_1-(\tilde{\alpha}\Lambda-\tilde{\beta}_x)\Phi_2,\\
\Psi_{1, t_2}=-(\tilde{r}\Lambda-\frac{1}{2}\tilde{s}_x)\Psi_1
+(\Lambda^{2}-\tilde{s}\Lambda-\frac{1}{2}\tilde{r}_x+\frac{1}{2}(\tilde{r}^{2}+\tilde{s}^{2})
+\tilde{\alpha}\tilde{\beta})\Psi_2+(\tilde{\beta}\Lambda+\tilde{\alpha}_x)\Psi_3,\\
\Psi_{2,
t_2}=-(\Lambda^{2}+\tilde{s}\Lambda+\frac{1}{2}\tilde{r}_x+\frac{1}{2}(\tilde{r}^{2}+\tilde{s}^{2})
+\tilde{\alpha}\tilde{\beta})\Psi_1+(\tilde{r}\Lambda-\frac{1}{2}\tilde{s}_x)\Psi_2
-(\tilde{\alpha}\Lambda-\tilde{\beta}_x)\Psi_3,\\
\Psi_{3,
t_2}=-(\tilde{\alpha}\Lambda-\tilde{\beta}_x)\Psi_1-(\tilde{\beta}\Lambda+\tilde{\alpha}_x)\Psi_2,
\end{array}\right.\end{equation}
where $\tilde{r}, \tilde{s}, \tilde{\alpha}, \tilde{\beta}$ denotes
the functions $r, s, \alpha, \beta$ defined by the explicit
constraint (\ref{D20}), and $\tilde{r}_x, \tilde{s}_x,
\tilde{\alpha}_x, \tilde{\beta}_x$ are given by
\begin{eqnarray*}\left\{\begin{array}{l}
\tilde{r}_x=2<\Lambda\Psi_1, \Phi_2>+2<\Lambda\Psi_2,
\Phi_1>+2<\Psi_1, \Phi_2>^{2}-2<\Psi_2, \Phi_1>^{2},\\
\tilde{s}_x=-2<\Lambda\Psi_1, \Phi_1>+2<\Lambda\Psi_2,
\Phi_2>+2(<\Psi_1, \Phi_1>-<\Psi_2, \Phi_2>)(<\Psi_2,
\Phi_1>-<\Psi_1, \Phi_2>),\\
\tilde{\alpha}_x=<\Lambda\Psi_1, \Phi_3>+<\Lambda\Psi_3, \Phi_2>
+(<\Psi_1, \Phi_2>-<\Psi_2, \Phi_1>)(<\Psi_1, \Phi_3>+<\Psi_3,
\Phi_2>),\\
\tilde{\beta}_x=<\Lambda\Psi_2, \Phi_3>-<\Lambda\Psi_3, \Phi_1>
-(<\Psi_1, \Phi_2>-<\Psi_2, \Phi_1>)(-<\Psi_2, \Phi_3>+<\Psi_3,
\Phi_1>),
\end{array}\right.\end{eqnarray*}
which are computed through using the spatial constrained flow
(\ref{D23}).

In what follows, we want to prove that the system (\ref{D21}) is a
completely integrable Hamiltonian system in the Liouville sense.
Furthermore, we shall prove that the system (\ref{D22}) is also
completely integrable under the control of the system (\ref{D21}).

On the one hand, the system (\ref{D21}) or (\ref{D23}) can be
represented as the following super Hamiltonian form
\begin{equation}\label{D25}
  \left\{\begin{array}{l} \displaystyle \Phi_{1, x}=\frac{\partial
H_1}{\partial\Psi_1}, \Phi_{2, x}=\frac{\partial
H_1}{\partial\Psi_2}, \Phi_{3, x}=\frac{\partial
H_1}{\partial\Psi_3},\vspace{2mm}
\\
\displaystyle \Psi_{1, x}=-\frac{\partial H_1}{\partial\Phi_1},
\Psi_{2, x}=-\frac{\partial H_1}{\partial\Phi_2}, \Psi_{3,
x}=\frac{\partial
H_1}{\partial\Phi_3},\end{array}\right.\end{equation} where
\begin{eqnarray*}
H_1&=&<\Lambda\Psi_1, \Phi_2>-<\Lambda\Psi_2, \Phi_1>+
\frac{1}{2}(<\Psi_1, \Phi_1>-<\Psi_2, \Phi_2>)^{2}
+\frac{1}{2}(<\Psi_1, \Phi_2>+<\Psi_2, \Phi_1>)^{2}\\&& +(-<\Psi_2,
\Phi_3>+<\Psi_3, \Phi_1>)(<\Psi_1, \Phi_3>+<\Psi_3, \Phi_2>).
\end{eqnarray*}
In addition, the characteristic property (\ref{D16}) and the
recurrence relations (\ref{D6}) ensure that
\begin{equation}\label{D26}\left\{\begin{array}{l}
\tilde{A}_{i+1}=<\Lambda^{i}\Psi_1, \Phi_2>+<\Lambda^{i}\Psi_2,
\Phi_1>, \quad i\geq0,\\
\tilde{B}_{i+1}=<\Lambda^{i}\Psi_2, \Phi_1>-<\Lambda^{i}\Psi_1,
\Phi_2>,\quad i\geq0,\\
\tilde{C}_{i+1}=<\Lambda^{i}\Psi_1, \Phi_1>-<\Lambda^{i}\Psi_2,
\Phi_2>,\quad i\geq0,\\
\tilde{\delta}_{i+1}=<\Lambda^{i}\Psi_1, \Phi_3>+<\Lambda^{i}\Psi_3,
\Phi_2>,\quad i\geq0,\\
\tilde{\rho}_{i+1}=-<\Lambda^{i}\Psi_2, \Phi_3>+<\Lambda^{i}\Psi_3,
\Phi_1>,\quad i\geq0.
\end{array}\right.\end{equation}
Then the co-adjoint representation equation $\tilde{V}_x=[\tilde{U},
\tilde{V}]$ remains true. Furthermore, we know that the equality
$\tilde{V}_x^{2}=[\tilde{U}, \tilde{V}^{2}]$ is also true. Let
\begin{equation}\label{D27}
F=\frac{1}{4}Str\tilde{V}^{2}.\end{equation} Then it is easy to find
that $F_x=0$, that is to say, F is a generating function of
integrals of motion for the system (\ref{D21}) or (\ref{D23}). Due
to $F=\sum_{n\geq0}F_n\lambda^{-n}$, we obtain the following
formulas of integrals of motion
\begin{equation}
F_0=-\frac{1}{2}\tilde{B}_0^{2},\quad F_1=-\tilde{B}_0\tilde{B}_1
,\quad
F_n=-\tilde{B}_0\tilde{B}_n+\frac{1}{2}\sum_{i=1}^{n-1}(\tilde{A}_i\tilde{A}_{n-i}-
\tilde{B}_i\tilde{B}_{n-i}+\tilde{C}_i\tilde{C}_{n-i}+2\tilde{\rho}_i\tilde{\delta}_{n-i}),
\quad n\geq2.\end{equation} Substituting (\ref{D26}) into the above
formulas of integrals of motion, we obtain the following expressions
of $F_m (m\geq0)$:
\begin{eqnarray}\label{D28}
F_0&=&-\frac{1}{2},\quad F_1=<\Psi_1, \Phi_2>-<\Psi_2, \Phi_1>,\nonumber\\
F_n&=&<\Lambda^{n-1}\Psi_1, \Phi_2>-<\Lambda^{n-1}\Psi_2, \Phi_1>+
\sum_{i=1}^{n-1}[2(<\Lambda^{i-1}\Psi_1,
\Phi_2><\Lambda^{n-i-1}\Psi_2,
\Phi_1>)\nonumber\\&&+\frac{1}{2}(<\Lambda^{i-1}\Psi_1,
\Phi_1>-<\Lambda^{i-1}\Psi_2, \Phi_2) (<\Lambda^{n-i-1}\Psi_1,
\Phi_1>-<\Lambda^{n-i-1}\Psi_2,
\Phi_2>)\nonumber\\&&+(-<\Lambda^{i-1}\Psi_2,
\Phi_3>+<\Lambda^{i-1}\Psi_3, \Phi_1>)(<\Lambda^{n-i-1}\Psi_1,
\Phi_3>+<\Lambda^{n-i-1}\Psi_3, \Phi_2)],\nonumber\\&&
n\geq2.\end{eqnarray}

On the other hand, let us consider the temporal part of the
nonlinearized system (\ref{D22}).  Making use of (\ref{D26}) and
(\ref{D28}), the system (\ref{D22}) can be represented as the
following super Hamiltonian form
\begin{equation}\label{D31`}\left\{\begin{array}{l} \displaystyle
\Phi_{1, t_n}=\frac{\partial F_{n+1}}{\partial\Psi_1}, \quad
\Phi_{2, t_n}=\frac{\partial F_{n+1}}{\partial\Psi_2}, \quad\Phi_{3,
t_n}=\frac{\partial F_{n+1}}{\partial\Psi_3}, \vspace{2mm}
\\
\displaystyle \Psi_{1, t_n}=-\frac{\partial
F_{n+1}}{\partial\Phi_1}, \quad \Psi_{2, t_n}=-\frac{\partial
F_{n+1}}{\partial\Phi_2}, \quad \Psi_{3, t_n}=\frac{\partial
F_{n+1}}{\partial\Phi_3}.
\end{array}\right.\end{equation}
This can be checked pretty easily. For example, we can show the last
but one equality in the above system as follows:
\begin{eqnarray*}
\Psi_{2,
t_n}&=&-\sum_{i=0}^{n}(\tilde{A}_i+\tilde{B}_i)\Lambda^{n-i}\Psi_1
+\sum_{i=0}^{n}\tilde{C}_i\Lambda^{n-i}\Psi_2
-\sum_{i=0}^{n}\tilde{\rho}_i\Lambda^{n-i}\Psi_3
\\&=&-\Lambda^{n}\Psi_1-2\sum_{i=1}^{n}<\Lambda^{i-1}\Psi_2,
\Phi_1>\Lambda^{n-i}\Psi_1 +\sum_{i=1}^{n}(<\Lambda^{i-1}\Psi_1,
\Phi_1>-<\Lambda^{i-1}\Psi_2, \Phi_2>)\Lambda^{n-i}\Psi_2
\\&&+\sum_{i=1}^{n}(<\Lambda^{i-1}\Psi_2, \Phi_3>-<\Lambda^{i-1}\Psi_3,
\Phi_1>)\Lambda^{n-i}\Psi_3\\
&=&-\frac{\partial F_{n+1}}{\partial\Phi_2}.\end{eqnarray*}

In order to further show the Liouville integrability for the
constrained flows (\ref{D21}) and (\ref{D22}), we need to prove the
commutative property of the integrals of motion $\{F_m\}_{m\ge 0}$,
under the corresponding Poisson bracket:
\begin{equation}\label{D30}
\{F, G\}=\sum_{i=1}^{3}\sum_{j=1}^{N}(\frac{\partial
F}{\partial\phi_{ij}}\frac{\partial
G}{\partial\psi_{ij}}-(-1)^{p(\phi_{ij})p(\psi_{ij})}\frac{\partial
F}{\partial\psi_{ij}}\frac{\partial
G}{\partial\phi_{ij}}).\end{equation} At this time, we still have an
equality $\tilde{V}_{t_n}=[\tilde{V}^{(n)}, \tilde{V}]$, and after a
similar discussion, we know that F is also a generating function of
integrals of motion for (\ref{D22}). Hence $F_m, \ m\geq0,$ are
integrals of motion for the system (\ref{D22}) or (\ref{D31`}),
which implies that

\begin{equation} \label{D31}
\{F_{m+1}, F_{n+1}\}=\frac{\partial}{\partial t_n}F_{m+1}=0, \quad
m, n\geq0. \end{equation} The above equality (\ref{D31}) shows that
$\{F_m\}_{m\geq0}$ are in involution in pair under the Poisson
bracket (\ref{D30}).

In addition, similarly to \cite{MaFO-PA1996}, we know that
\begin{equation}\label{D32}
f_k=\psi_{1k}\phi_{1k}+\psi_{2k}\phi_{2k}+\psi_{3k}\phi_{3k},\quad
1\leq k\leq N,\end{equation} are integrals of motion for (\ref{D21})
and (\ref{D22}). It is not difficult to verify that 3N functions
$\{f_k\}_{k=1}^{N}$, $\{F_m\}_{m=1}^{2N}$ are in involution in pair.
To show the functional independence of 3N functions
$\{f_k\}_{k=1}^{N}$, $\{F_m\}_{m=1}^{2N}$, we can use, as in
reference \cite{HYZC}, the technique developed by Ma et al. in
\cite{MaFO-PA1996, MaZ-JMP2001}. Therefore, 3N functions
$\{f_k\}_{k=1}^{N}$, $\{F_m\}_{m=1}^{2N}$ are functionally
independent over some region of the supersymmetry manifold
$R^{4N|2N}$. Now, all of the above analysis gives the following
theorem.
\begin{Theorem}\label{superdiracthm1}
Both the spatial and temporal constrained flows (\ref{D21}) and
(\ref{D22}) are Liouville integrable super Hamiltonian systems
defined on the supersymmetry manifold $R^{4N|2N}$, which possess
{\rm 3N} functionally independent and involutive integrals of motion
$\{f_k\}_{k=1}^{N}$ and $\{F_m\}_{m=1}^{2N}$ defined by (\ref{D32})
and (\ref{D28}). Moreover, the formula (\ref{D20}) provides a
B\"acklund transform from the constrained flows (\ref{D21}) and
(\ref{D22}) to the Dirac systems (\ref{D12}).
\end{Theorem}

\begin{Remark}\label{Rem2}
The super-system on supermanifolds $R^{2M|2N}$ is Liouville
integrable \cite{jetzer} if it possesses $M$ even valued conserved
quantities and $N$ odd valued conserved quantities that are
independent and are also in involution.  Furthermore, similar to the
classical case, there exist a super analogue of Liouville's theorem
\cite{jetzer,shander}.  Note that, set $\theta$ be a odd variable in superspace,
 then $\tilde{f}_k=\theta f_k$,$k=1,2,\cdots,N$, are $N$ odd valued conserved quantities
 for super-finite dimensional in theorem \ref{superdiracthm1} because this system is involved only with
 even flow $t_n$.
\end{Remark}

\section{Conclusions and remarks}

In this paper, we have applied the binary nonlinearization method to the super Dirac systems by the Bargmann symmetry constraint
(\ref{D20}). We have also shown in Theorem \ref{superdiracthm1} that the nonlinearized systems (\ref{D21}) and (\ref{D22}) are two
super finite-dimensional integrable Hamiltonian systems, whose super Hamiltonian forms and integrals of motion have been presented
explicitly. We would also like to emphasize that the new formula (\ref{L1})  is a general result for calculating the variational
derivative of the spectral parameter $\lambda $ with respect to the  potential $u$. The crucial difference between the nonlinearziation
processes of the super AKNS system and the super Dirac system is due to the variational derivatives of $\lambda $ defined by formula (21)
in reference \cite{HYZC} and formula (\ref{D15}) in this paper.

We remark that the super Dirac systems (\ref{D9}) or (\ref{D12}),
e.g., (\ref{D13}), only possess super (odd and even) independent
variables. The fully supersymmetric Dirac systems possessing both
super dependent variables and super independent variables seem to be
a very interesting object for our future research. For more detailed
discussions on the supersymmetry theory and supersymmetric analysis,
we would like to refer readers to reference \cite{cartier}.

\vskip 2mm

{\bf Acknowledgments} {\small This work is supported by the Hangdian
Foundation KYS075608072, and NSF of China under Grant No. 10671187.
Jingsong He is also supported by Program for NCET under Grant
No.NCET-08-0515. We thank professor Yishen Li (USTC, China) for
valuable discussions on this topic. }

\end{document}